\documentclass[12pt]{article}
\usepackage{multirow,amsmath,amssymb,amsthm,graphicx,geometry,setspace,xr}
\usepackage[authoryear,longnamesfirst]{natbib}
\usepackage[hidelinks]{hyperref}
\newtheorem{theorem}{Theorem}

\newtheorem{assumption}{Assumption}
\DeclareMathOperator*{\argmin}{arg\,min}
\allowdisplaybreaks

\title{{\Large Unconditional Quantile Regression with High-Dimensional Data}\thanks{The first arXiv date:  July 27, 2020. We thank Layal Lettry for useful advice on the Job Corps survey data. We would like to thank Colin Cameron, Andres Santos (Co-Editor), and anonymous referees for very helpful comments. We also benefited from very useful comments by participants in the seminar at University of New Hampshire, University of Surrey, Syracuse University, the Presidential Session at the SEA 90th Annual Meeting, 2021 North American Summer Meeting of Econometric Society, and 2022 North American Winter Meeting of Econometric Society. Yichong Zhang acknowledges the financial support from Singapore Ministry of Education Tier 2 grant under grant MOE2018-T2-2-169 and the Lee Kong Chian fellowship. The usual disclaimer applies.}}
\author{Yuya Sasaki\thanks{Y. Sasaki: Department of Economics, Vanderbilt University, VU Station B \#351819, 2301 Vanderbilt Place, Nashville, TN 37235-1819. Email: yuya.sasaki@vanderbilt.edu} \and Takuya Ura\thanks{T. Ura: Department of Economics, University of California, Davis, One Shields Avenue, Davis, CA 95616. Email: takura@ucdavis.edu}  \and Yichong Zhang\thanks{Y. Zhang: School of Economics, Singapore Management University, 90 Stamford Road, Singapore 178903, Singapore. Email: yczhang@smu.edu.sg}}
\date{}
\begin{document}
\maketitle
\begin{abstract}
This paper considers estimation and inference for heterogeneous counterfactual effects with high-dimensional data. We propose a novel robust score for debiased estimation of the unconditional quantile regression \citep{firpo2009unconditional} as a measure of heterogeneous counterfactual marginal effects. We propose a multiplier bootstrap inference and develop asymptotic theories to guarantee the size control in large sample. Simulation studies support our theories. Applying the proposed method to Job Corps survey data, we find that a policy which counterfactually extends the duration of exposures to the Job Corps training program will be effective especially for the targeted subpopulations of lower potential wage earners.
\begin{description}
\item[Keywords:] counterfactual analysis, debiased machine learning, doubly/locally robust score
\end{description}
\end{abstract}

\newpage
\section{Introduction}
Analysis of an outcome response to a counterfactual shift in the covariate distribution is of interest in policy studies. Such a counterfactual analysis requires accounting for the Oaxaca-Blinder decomposition of heterogeneous outcome distributions into structural heterogeneity ($F_{Y|X}$) and distributional heterogeneity ($F_X$); see \citet{fortin2011decomposition} for a review. To conduct a credible counterfactual analysis, it is crucial to control a structure ($F_{Y|X}$) with rich information about $X$ while applying a counterfactual shift in the distribution of $X$. In this light, a researcher ideally wants to use high-dimensional $X$ in data.

Motivated by this feature of causal inference and the recently increasing availability of high-dimensional data, we develop a novel theory and method of estimation and inference for heterogeneous counterfactual effects with high-dimensional controls. The existing literature features a number of alternative approaches and frameworks of counterfactual analysis. Among others, we focus on the unconditional quantile partial effect \citep*[UQPE;][]{firpo2009unconditional} in the unconditional quantile regression based on the re-centered influence function (RIF) of \citet{firpo2009unconditional} for two reasons: (i) its advantage of providing ``a simple way of performing detailed decompositions'' \citep[][p. 76]{fortin2011decomposition} and (ii) its popularity.\footnote{As of February 17, 2022, \citet{firpo2009unconditional} have attracted 2275 Google Scholar citations.} This parameter measures the marginal effect of counterfactually shifting the distribution of a coordinate of $X$ on population quantiles of an outcome.

The UQPE is defined with the conditional distribution $F_{Y|X}$ and the marginal distribution $F_X$. Let $X=(X_1,X_{-1})$ denote the status quo, where $X_1$ is a scalar treatment variable of interest and $X_{-1}$ consists of controls. We focus on the change from $X$ to $(X_1+\varepsilon,X_{-1})$ throughout this paper, while our analysis can be generalized to the change in any fixed direction. The counterfactual distribution of $Y$ after this change is 
$$
F_{Y}^\varepsilon(y)=\int F_{Y|X=(x_1+\varepsilon,x_{-1})}(y)dF_X(x).
$$
The UQPE with respect to the first coordinate, $X_1$, of $X$ is defined by 
\begin{equation}\label{eq:parameterofinterest}
UQPE(\tau)=\left.\frac{\partial Q_\tau(F_{Y}^\varepsilon)}{\partial\varepsilon}\right|_{\varepsilon=0},
\end{equation}
where $Q_\tau$ is the $\tau$-th quantile operator. The UQPE measures the change in the outcome quantile when the distribution of $X$  changes infinitesimally in the direction of the first coordinate. Under the assumption of conditional exogeneity, as in \citet[Section 2.3]{chernozhukov2013inference}, the UQPE can be interpreted as the causal effect of changing the distribution of $X$ infinitesimally. Without such an assumption, UQPE may still be of interest as a summary statistic of the counterfactual distributional relationship between $Y$ and $X_1$.

To fix ideas, suppose that a policy maker is interested in analyzing the counterfactual effects of extending the duration $X_1$ of an exposure to the Job Corps training program on the outcome $Y$ of hourly wages, controlling for a large number of demographic, socioeconomic and behavioral attributes $X_{-1}$. The median of the actual distribution of hourly wages is given by $Q_{0.5}(F_Y^0)$. On the other hand, if the exposures are extended by $\varepsilon$ days for every participant, then the median of the counterfactual distribution of hourly wages becomes $Q_{0.5}(F_Y^\varepsilon)$. In this case, $Q_{0.5}(F_Y^\varepsilon) - Q_{0.5}(F_Y^0)$ measures the counterfactual effect on the median of the wage distribution, and $UQPE(\tau) = \lim_{\varepsilon \rightarrow 0} \big(Q_{0.5}(F_Y^\varepsilon) - Q_{0.5}(F_Y^0)\big)/\varepsilon$ measures its marginal effect.
 
While the RIF regression approach is indeed simpler to implement than alternative methods of counterfactual analysis as emphasized by \citet{fortin2011decomposition}, an estimation of the UQPE still requires a three-step procedure. The first step is an estimation of unconditional quantiles. The second step implements the RIF regression. The third step integrates the RIF regression estimates to in turn estimate the UQPE. \citet[]{firpo2009unconditional} provide an estimation procedure for the case of low-dimensional data. If we allow for high-dimensional controls with the aforementioned motivation, then the second step will require some estimation of the high-dimensional RIF regression, and the traditional techniques to incorporate estimation errors of the second step into the third step no longer apply. To overcome this challenge, we construct a novel doubly/locally robust score for estimation of the UQPE. The key insight for the construction is the identification result in \citet[p. 958]{firpo2009unconditional} that the UQPE has the same structure as the average derivative estimator, whose influence function  in the presence of nonparametric preliminary estimation has been well studied in the existing literature (e.g., \citealp{newey1994asymptotic}). With this doubly/locally robust score, we obtain a Z-estimation criterion with robustness against perturbations in functional nuisance parameters as in \cite{belloni2014uniform} and \cite{belloni2018uniformly}, and can thereby use the debiased estimation approach \citep[e.g.,][]{belloni2014uniform,chernozhukov2017double,chernozhukov2018double,chernozhukov2016locally}, which allows one to obtain the asymptotic distribution of a UQPE estimator, independently of the second-step estimation as far as it satisfies some convergence rate conditions satisfied by major nonparametric estimators and machine learners. To provide a readily applicable method for practitioners, we focus on a specific method of estimation and bootstrap inference in the main text, but we also provide a generic method in the online supplement.

\noindent\textbf{Notations.} In this paper, we will use the following mathematical symbols and notations. 
$\mathcal{X}$ denotes the support of $X$. 
For a vector $v$, we define $\mathrm{Supp}(v)=\{i: v_i\ne 0\}$,  $\|v\|_1=\sum_{i}|v_i|$, $\|v\|_2=(\sum_{i}v_i^2)^{1/2}$, and $\|v\|_\infty=\max_{i}|v_i|$.
Denote the cardinality of $\mathrm{Supp}(v)$ by $\|v\|_0$.
For a matrix $A$, we define $\|A\|_\infty = \max_{ij}|A_{ij}|$.
We let $\Lambda$ be the standard logistic CDF and $\Phi$ be the standard normal CDF.

\section{Robust Score, Estimation, and Inference for $UQPE(\tau)$}\label{sec:scoreconstruction}
In this section, we develop a new score for doubly/locally robust estimation of the UQPE. We then present specific estimation and inference procedures in Sections \ref{sec:est_procedure} and \ref{sec:inf_procedure}, respectively. It is worthwhile to mention here that our analysis allows the dimensionality of $X$ to depend on the sample size $N$ and to diverge as $N\rightarrow\infty$. 

Following \citet{firpo2009unconditional}, we can rewrite our parameter of interest, defined in \eqref{eq:parameterofinterest}, as a function of identifiable objects. Namely,  
$$
UQPE(\tau)=-\frac{\theta(\tau)}{f_Y(q_\tau)},
$$
where $f_Y$ is the density function of $Y$, $q_\tau$ is the $\tau$-th quantile of $Y$, and 
\begin{equation}\label{eq:definition_theta} 
\theta(\tau)=\int \frac{\partial F_{Y\mid X=x}(q_\tau)}{\partial  x_1}dF_X(x).
\end{equation}
This equation is shown in \citet[Corollary 1]{firpo2009unconditional}.

\subsection{Doubly/Locally Robust Score}
We could estimate $\theta(\tau)$ based on \eqref{eq:definition_theta} and some estimator for $F_{Y\mid X}(\cdot)$. When $X$ is high-dimensional, this direct estimation of $\theta(\tau)$ can result in a large bias, a large variance, or both. Instead, we propose to construct an estimator for $\theta(\tau)$ based on an alternative representation:
\begin{align}
{\theta}(\tau)
&=
\int \frac{\partial F_{Y\mid X=x}(q_\tau)}{\partial  x_1}dF_X(x)-\int 
\omega(x)(1\{y\leq q_\tau\}-{m}_0(x,q_\tau))
dF_{Y,X}(y,x)\notag\\
&=
\int \Big(
{m}_1(x,q_\tau)-\omega(x)\big(1\{y\leq q_\tau\}-{m}_0(x,q_\tau)\big)
\Big)
dF_{Y,X}(y,x),
\label{eq:moment_conditions}
\end{align}
where $\omega(x)=\partial\log f_{X_1|X_{-1}=x_{-1}}(x_1)/\partial x_1$, $m_0(x,q)=F_{Y\mid X=x}(q)$ and $m_1(x,q)={\partial m_0(x,q)}/{\partial  x_1}$. This representation in \eqref{eq:moment_conditions} comes from the influence adjustment term for the average derivative estimator \citep[][p.1369]{newey1994asymptotic}. Namely, $\int \omega(x)(1\{y\leq q_\tau\}-{m}_0(x,q_\tau))dF_{Y,X}(y,x)$ in \eqref{eq:moment_conditions} adjusts the estimation error from the regularized preliminary estimation.  

The advantage of \eqref{eq:moment_conditions} over \eqref{eq:definition_theta} is that \eqref{eq:moment_conditions} is doubly robust in the sense that
\begin{equation}\label{eq:doublerobust1}
\theta(\tau)=\int\left(\tilde{m}_1(x,q_\tau)-\omega(x)(1\{y\leq q_\tau\}-\tilde{m}_0(x,q_\tau))\right)dF_{Y,X}(y,x)
\end{equation}
and 
\begin{equation}\label{eq:doublerobust2}
\theta(\tau)=\int\left({m}_1(x,q_\tau)-\tilde\omega(x)(1\{y\leq q_\tau\}-{m}_0(x,q_\tau))\right)dF_{Y,X}(y,x)
\end{equation}
hold for a set of values that the high-dimensional nuisance parameters $(\tilde\omega(x),$ $\tilde{m}_{0}(x,q),$ $\tilde{m}_{1}(x,q))$ take as far as $\tilde{m}_1(x,q)={\partial\tilde{m}_0(x,q)}/{\partial  x_1}$ and some regularity conditions to be formally stated below are satisfied. Note that $(\tilde\omega(x),\tilde{m}_{0}(x,q),\tilde{m}_{1}(x,q))$ in \eqref{eq:doublerobust1} and \eqref {eq:doublerobust2} can be different from the true value $(\omega(x),{m}_{0}(x,q),{m}_{1}(x,q))$. Our doubly/locally robust moment involves two nuisance parameters, $\omega(x)$ and $m_0(x,q)$, which are based on the conditional distribution of $X_1$ on $X_{-1}$ and the conditional distribution of $Y$ on $X$, respectively. They are analogous to the propensity score and the conditional mean function in the doubly robust moment for the average treatment effect. The equality in \eqref{eq:doublerobust1} (resp. \eqref{eq:doublerobust2}) implies that, even if we mis-specify the conditional distribution of $Y$ given $X$ (resp. the conditional distribution of $X_1$ given $X_{-1}$), \eqref{eq:moment_conditions} provides the correct parameter value  $\theta(\tau)$.  The construction of our doubly/locally robust moment comes from the fact that  ${\theta}(\tau)$ can be represented in two different ways: 
$$
{\theta}(\tau)=\int {m}_1(x,q_\tau)dF_X(x)
\quad\mbox{ and }\quad
{\theta}(\tau)=-\int \omega(x)1\{y\leq q_\tau\}dF_{Y,X}(y,x).
$$
Each of the two representations only requires one of the two functions, ${m}_1(x,q_\tau)$ and $\omega(x)$, to be specified correctly. A precise statement for the double robustness and its proof are found in Appendix \ref{sec:doublerobust} in the online supplement.

\subsection{Estimation Procedure}
\label{sec:est_procedure}
With the sample $\{(Y_i,X_i): i=1,\ldots,N\}$ and the moment condition \eqref{eq:moment_conditions}, we propose to estimate ${\theta}(\tau)$ by a plug-in method. Let $(\hat\omega(x),\hat{m}_{0}(x,q),\hat{m}_{1}(x,q))$ denote an estimator of $(\omega(x),{m}_{0}(x,q),{m}_{1}(x,q))$ -- a concrete procedure to construct  $(\hat\omega(x),\hat{m}_{0}(x,q),\hat{m}_{1}(x,q))$ is provided below. Letting $\hat{q}_\tau$ denote the sample $\tau$-th empirical quantile of $Y$, we estimate $\theta(\tau)$ by 
\begin{align}
\label{eq:thetahat}
\hat{\theta}(\tau)=\frac{1}{N}\sum_{i=1}^N \left(\hat{m}_{1}(X_i,\hat{q}_\tau) - \hat\omega(X_i)(1\{Y_i \leq \hat{q}_\tau\} - \hat{m}_{0}(X_i,\hat{q}_\tau))\right).
\end{align}
With this estimator for $\theta(\tau)$, our proposed estimator for $UQPE(\tau)$ is in turn defined by
$$
\widehat{UQPE}(\tau) = -\frac{\hat{ \theta}(\tau)}{\hat{f}_Y(\hat{q}_\tau)},
$$
where  $\hat{f}_Y(y)$ is the kernel density estimator defined by
$$
\hat{f}_Y(y) = \frac{1}{N}\sum_{i=1}^N\frac{1}{h_1}K_1\left(\frac{Y_i - y}{h_1}\right)
$$ 
for a kernel function $K_1$ and a bandwidth parameter $h_1$.

We use the logistic Lasso regression \citep{BCFH13} to construct $\hat{m}_{0}(x,q)$. Once we construct $\hat{m}_{0}(x,q)$, we in turn define $\hat{m}_{1}(x,q)$ by 
$$
\hat{m}_{1}(x,q) = \frac{\partial\hat{m}_{0}(x,q)}{\partial x_1}. 
$$
Consider the approximately sparse logistic regression model for  $m_0(x,q)$:
$$
m_0(X,q) = \Lambda(b(X)^\top\beta_{q})+(\mathrm{approximation\ error}),
$$ 
where $b(X)$ is a $p_b$-dimensional observed vector and $\beta_{q}$ is an unknown parameter. Assumption \ref{assn:Lasso} (to be stated below in Section \ref{sec:theoretical_result}) specifies the conditions for the sparsity and formalizes the approximation error. In our numerical examples, we define $b(X)$ by including powers of $X$ up to the third degree and standardize each component of $b(X)$ so that the variance is one. We estimate $\beta_{q}$ by the Lasso penalized logistic regression 
\begin{equation}\label{eq:m0}
\tilde{\beta}_{q} = \argmin_\beta -\frac{1}{N}\sum_{i=1}^N\log\left(\Lambda(b(X_i)^\top\beta)^{1\{Y_{i} \leq q\}}(1-\Lambda(b(X_i)^\top\beta))^{1\{Y_{i}>q\}}\right)+ \frac{\lambda_{L}}{N}\|{\Psi}_q\beta\|_1,
\end{equation}
where ${\Psi}_q$ is a diagonal matrix with penalty loadings defined in the next paragraph. We follow \cite{BCFH13} and set the regularization parameter as 
$$
\lambda_{L} = 1.1\Phi^{-1}(1-(0.1/\log(N))/(p_b\vee N))N^{1/2}.
$$
We recommend using the post-Lasso estimator for $\beta_q$ defined by 
$$
\hat{\beta}_{q} = \argmin_{\beta \in \mathbb{R}^p: \mathrm{Supp}(\beta)\subset(\mathrm{Supp}(\tilde{\beta}_{q}) \cup S_1)}-\frac{1}{N}\sum_{i=1}^N\log\left(\Lambda(b(X_i)^\top\beta)^{1\{Y_{i} \leq q\}}(1-\Lambda(b(X_i)^\top\beta))^{1\{Y_{i}>q\}}\right),
$$
where $S_1 \subset \{1,\ldots,p_b\}$ denotes the coordinate set of covariates researchers want to include in the post-Lasso regression. 
For the UQPE with respect to $X_1$, it is natural to include $X_1$ in the regression. The post-Lasso estimator can do so by setting $1\in S_1$, whereas the Lasso estimator $\tilde{\beta}_{q}$ may exclude $X_1$ from the regression.
With $\hat{\beta}_{q}$, we can estimate $\hat{m}_{0}(x,q)$ by
$$
\hat{m}_{0}(x,q) = \Lambda(b(x)^\top \hat{\beta}_{q}). 
$$

The penalty loading matrix ${\Psi}_q=\text{diag}(\psi_{q,1},\cdots ,\psi_{q,p_b})$ needs to be estimated to implement \eqref{eq:m0}. Ideally, we would like to use the infeasible penalty loading   
$$
\bar{\psi}_{q,j}=\sqrt{\frac{1}{N}\sum_{i=1}^N\left(1\{Y_i\leq q\}-m_{0}(X_i,q)\right)^2b^2_{j}(X_i)}.
$$ 
Since $m_{0}(X,q)$ is unknown, \cite{BCFH13} propose the following iterative algorithm to obtain the feasible version of the loading matrix:
\begin{enumerate}
\item We start the algorithm with $\psi_{q,j}^{0}=\sqrt{\frac{1}{N}\sum_{i=1}^N1\{Y_i \leq q\}b^2_{j}(X_i)}.$ 
\item For $k=0,\cdots ,K-1$ for some fixed positive integer $K$, 
we can compute $\tilde{\beta}_{q}^{k}$ by \eqref{eq:m0} with $\tilde{\Psi}_{q}^{k}=\text{diag}(\psi_{q,1}^{k},\cdots,\psi_{q,p_b}^{k})$, and construct 
$$
\psi_{q,j}^{k+1}=\sqrt{\frac{1}{N}\sum_{i=1}^N \left(1\{Y_i\leq q\}-\Lambda (b(X_i)^\top\tilde{\beta}_{q}^{k})\right)^2b^2_{j}(X_i)}.
$$
\item The final penalty loading matrix ${\Psi}_{q}^K=\text{diag}(\psi_{q,1}^K,\cdots,\psi_{q,p_b}^K)$ will be used for ${\Psi}_{q}$ in \eqref{eq:m0}. 
\end{enumerate}

Next, we consider a regularized estimation of $\omega(X)$ based on the Riesz representer approach \citep{chernozhukov2018automatic,chernozhukov2018double2}. Suppose that $h(x)$ is a $p_h$-dimensional dictionary of approximating functions that are differentiable in $x_1$ and that   
$$
\omega(x)=h(x)^\top\overline{\rho}+(\mathrm{approximation\ error})
$$
holds, where Assumption \ref{ass:rr} (to be stated below in Section \ref{sec:theoretical_result}) formally describes this  approximation. 
In our numerical examples, we define $h(X)$ by including powers of $X$ up to the third degree and standardize each component of $h(X)$ so that the variance is one. Since $\omega(x)={\partial \log f_{X_1|X_{-1}=x_{-1}}(x_1)}/{\partial x_1}$, the integration by parts yields 
$$
\mathbb{E}[h(X) \omega(X)] = - \mathbb{E}[\partial_{x_1}h(X)].
$$
Approximating $\omega(x)$ by $h(x)^\top\overline{\rho}$, we have 
$$
\mathbb{E}[h(X)h(X)^\top]\overline{\rho}=- \mathbb{E}[\partial_{x_1}h(X)]+(\mathrm{approximation\ error}).
$$
Thus, $\overline{\rho}$ can be approximated by $\argmin_{\rho}\left(-2M^\top \rho + \rho^\top G\rho\right)$, where $G = \mathbb{E}[h(X)h(X)^\top]$ and $M = -\mathbb{E}[\partial_{x_1}h(X)]$. To accommodate high-dimensional $h(x)$, we use the regularized minimizer
$$
\argmin_{\rho}\left(-2M^\top \rho + \rho^\top G\rho +  \lambda_{R} \|\rho\|_1\right)
$$
with $\lambda_{R}$ denoting a regularization parameter (cf. Assumption \ref{ass:rr}.5). In the simulations and empirical application, we use $\lambda_{R} = 2\log(\log(N)) \sqrt{\log(p_h)/N}$. The Riesz representer approach uses the sample analog of this objective to estimate $\omega(x)$. Namely, we estimate $\omega(x)$ by $$\hat\omega(x) = h(x)^\top \hat{\rho},$$ where $\hat{G} = \frac{1}{N}\sum_{i=1}^Nh(X_i)h(X_i)^\top$,  $\hat{M} = -\frac{1}{N}\sum_{i=1}^N\partial_{x_1}h(X_i)$, and
$$
\hat{\rho} = \argmin_{\rho}\left(-2 \hat{M}^\top \rho + \rho^\top \hat{G} \rho +  \lambda_{R} \|\rho\|_1\right).
$$

\subsection{Bootstrap Inference}\label{sec:inf_procedure}
For an inference about $UQPE(\tau)$, we propose the multiplier bootstrap without requiring to recalculate the preliminary estimator $\hat\omega(x)$ in each bootstrap iteration. (More precisely, if we calculate $(\hat m_0(x,q), \hat m_1(x,q))$ on a grid of values of $q$ once, then we do not need to recalculate them in each bootstrap iteration either.) Using independent standard normal random variables $\{\eta_i\}_{i=1}^N$ that are independent of the data, we compute the bootstrap estimators $\hat{ \theta}^*(\tau)$ and $\widehat{UQPE}^*(\tau)$ in the following steps. The bootstrap estimator for $q_\tau$ is $\hat{q}^*_\tau$ defined by the $r_N^*$-th order statistic of $Y_i$, where $r_N^*$ is the integer part of $1+\sum_{i=1}^N\left(\tau+\eta_i(\tau - \mathbf{1}\{Y_i \leq \hat{q}_\tau\})\right)$.\footnote{It is equivalent to $\hat{q}^*_\tau = \arg\min_q \sum_{i=1}^N \rho_\tau(Y_i - q) - q \sum_{i=1}^N \eta_i(\tau - 1\{Y_i \leq \hat{q}_\tau\})$, which is the gradient bootstrap method \citep{chen2004quantile} and directly perturbs the score for the quantile $q_\tau$. By the sub-gradient condition, we have that $\hat{q}^*_\tau$ equals the $r_N^*$th order statistic of $Y_i$, where  $r_N^*$ is the integer that satisfies $N\tau + \sum_{i=1}^N \eta_i(\tau - 1\{Y_i \leq \hat{q}_\tau\}) + 1 \geq r_N^* \geq N\tau + \sum_{i=1}^N \eta_i(\tau - 1\{Y_i \leq \hat{q}_\tau\})$. This procedure gives us a closed-form solution for $\hat{q}^*_\tau$.}
The bootstrap estimators for $f_Y(y)$ and $\theta(\tau)$ are 		 
$$
\hat{f}_Y^*(y) = \frac{1}{\sum_{i=1}^N(\eta_i+1)}\sum_{i=1}^N(\eta_i+1)\frac{1}{h_1}K_1\left(\frac{Y_i - y}{h_1}\right), 
$$
and 
$$
\hat{\theta}^*(\tau)= \frac{1}{\sum_{i=1}^N(\eta_i+1)}\sum_{i=1}^N (\eta_i+1)\left(\hat{m}_{1}(X_i,\hat{q}^*_\tau) - \hat\omega(X_i)(1\{Y_i \leq \hat{q}^*_\tau\} - \hat{m}_{0}(X_i,\hat{q}^*_\tau))\right),
$$
respectively. With these components, the bootstrap estimator $\widehat{UQPE}^*(\tau)$ is given by
$$
\widehat{UQPE}^*(\tau) = -\frac{\hat{ \theta}^*(\tau)}{\hat{f}^*_Y(\hat{q}^*_\tau)}.
$$

We can use the above multiplier bootstrap method to conduct various types of statistical inference about the UQPE. First, testing the hypothesis of  $UQPE(\tau) = 0, \forall\tau\in \Upsilon$ for some closed interval $\Upsilon\subset(0,1)$ is of main interest in many empirical applications. Because $f_Y(q_\tau)$ is assumed to be bounded away from zero, such a hypothesis is equivalent to $\theta(\tau) = 0, \forall\tau \in \Upsilon$, where $\theta(\tau)$ can be estimated in a parametric rate. We can thus test $UQPE(\tau) = 0,  \forall\tau\in \Upsilon$ by constructing a confidence band for $\{\theta(\tau): \tau \in \Upsilon  \}$ and checking whether the constant zero function belongs to this band. Specifically, let
$$
\hat{\sigma}^{\theta}(\tau) = \frac{Q_{\hat\theta^*(\tau)}(0.75) - Q_{\hat\theta^*(\tau)}(0.25)}{\Phi^{-1}(0.75)-\Phi^{-1}(0.25)}
$$
denote an estimator of the standard error of $\hat{\theta}(\tau)$ for $\tau \in \Upsilon$, where $Q_{\hat\theta^*(\tau)}(0.75)$ and $Q_{\hat\theta^*(\tau)}(0.25)$ denote the 75th and 25th percentiles of $\hat\theta^*(\tau)$ conditional on the data. Let $c_{\Upsilon}^{\theta}(1-\alpha)$ denote the $(1-\alpha)$ quantile of 
$$
\sup_{\tau \in \Upsilon} \left\vert \frac{\hat{\theta}^*(\tau)-\hat{\theta}(\tau)}{\hat\sigma^{\theta}(\tau)} \right\vert
$$ 
conditional on the data. Let $CB^{\theta}_{\Upsilon}$ denote the confidence band of $\theta(\cdot)$ on $\Upsilon$ whose lower and upper bounds at $\tau \in \Upsilon$ are given by $\hat{\theta}(\tau) \pm  \hat\sigma^{\theta}(\tau) c_{\Upsilon}^\theta(1-\alpha)$. 

Second, we can similarly construct a confidence band for $\{UQPE(\tau):\tau\in \Upsilon\}$. Let
$$
\hat\sigma(\tau) = \frac{Q_{\widehat{UQPE}^*(\tau)}(0.75) - Q_{\widehat{UQPE}^*(\tau)}(0.25)}{\Phi^{-1}(0.75)-\Phi^{-1}(0.25)}
$$ 
denote an estimator of the standard error of $\widehat{UQPE}(\tau)$ for $\tau \in \Upsilon$, where $Q_{\widehat{UQPE}^*(\tau)}(0.75)$ and $Q_{\widehat{UQPE}^*(\tau)}(0.25)$ denote the 75th and 25th percentiles of $\widehat{UQPE}^*(\tau)$ conditional on the data. Let $c_\Upsilon(1-\alpha)$ denote the $(1-\alpha)$ quantile of 
$$
\sup_{\tau \in \Upsilon} \left\vert \frac{\widehat{UQPE}^*(\tau)-\widehat{UQPE}(\tau)}{\hat\sigma(\tau)}\right\vert
$$ 
conditional on the data. Let $CB_\Upsilon$ denote the confidence band of $UQPE(\cdot)$ on $\Upsilon$ whose lower and upper bounds at $\tau \in \Upsilon$ are given by $\widehat{UQPE}(\tau) \pm \hat\sigma(\tau) c_\Upsilon(1-\alpha)$.

\section{Asymptotic Theory}\label{sec:theoretical_result}
In this section, we investigate the asymptotic properties of the estimators $(\hat{ \theta},\widehat{UQPE})$ and their bootstrap counterparts $(\hat{ \theta}^*,\widehat{UQPE}^*)$ introduced in the previous section. The uniformity over $\tau$ is relevant to applications (e.g., analysis of heterogeneous counterfactual effects across $\tau$), and therefore, in this section, we aim to control the residuals for the linear expansion uniformly over $\tau\in \Upsilon$ for some closed interval $\Upsilon\subset(0,1)$. Let $\mathcal{Q} = \{{q}_\tau: \tau \in \Upsilon\}$, and let $\mathcal{Q}^\delta$ denote the $\delta$ enlargement of $\mathcal{Q}$.

\begin{assumption}
\begin{enumerate}
\item[]
\item For every $\tau \in \Upsilon$, $F_{Y}^\varepsilon(q)$ is differentiable with respect to $\varepsilon$ in a neighborhood of zero for every $q$ in a neighborhood of $q_\tau$, and $Q_\tau(F_{Y}^\varepsilon)$ is well defined and is differentiable with respect to $\varepsilon$ in a neighborhood of zero. 
\item $\int\left(\sup_{q \in \mathcal{Q}^\delta}|{m}_1(x,q)|\right)^{2+d}dF_X(x)$ and $\int|\omega(x)|^{2+d}dF_X(x)$ are finite for some $d>0$.
\item  For every $x_{-1}$ in the support of $X_{-1}$, the conditional distribution of $X_1$ given $X_{-1}=x_{-1}$ has a probability density function, denoted by $f_{X_1\mid X_{-1}}$, which is continuously differentiable everywhere and is zero on the boundary of the support of the conditional distribution of $X_1$. 
\item  $m_1(x,q)$ and $m_0(x,q)$ are differentiable with respect to $q$ for $q \in \mathcal{Q}^\delta$, and the derivatives are bounded in absolute value uniformly over $x \in \mathcal{X}$ and $q \in \mathcal{Q}^\delta$. 
\item $f_Y(y)$ is three times differentiable on $\mathcal{Q}^\delta$ with all the derivatives uniformly bounded. $f_Y(q_\tau)>0$ for every $\tau \in \Upsilon$. 
\end{enumerate}
\label{assn:identification}
\end{assumption}

This assumption is on the model primitives.  Assumptions \ref{assn:identification}.1 and \ref{assn:identification}.3--\ref{assn:identification}.6 impose regularity in terms of the smoothness of various functions representing features of the data.  We use the enlargement $\mathcal{Q}^\delta$ instead of $\mathcal{Q}$ because with probability approaching one, $\hat{q}_\tau$ and $\hat{q}^*_\tau$ belong to the former but not necessarily the latter. Assumption  \ref{assn:identification}.2 is the standard moment condition. 

We impose the following condition to bound the estimation error for $\hat{m}_{j}(x,q)$. 

\begin{assumption}
	\begin{enumerate}
		\item[]
\item\label{assn:Lasso_boundedmoments}  (Boundedness)
		For some positive constants $\delta,\overline{c},\underline{c}$,
(i) $
			\underline{c} \leq \int b_j(x)^2dF_X(x) \leq \overline{c}\mbox{ for every }j=1,\ldots,p,
			$
			(ii) $
			\sup_{x \in \mathcal{X},q\in \mathcal{Q}^\delta}|m_1(x,q)| \leq \overline{c},
			$ and 
			(iii) $
			\sup_{x \in \mathcal{X},q\in \mathcal{Q}^\delta}|\frac{\partial}{\partial x_1}b(x)^\top\beta_q| \leq \overline{c}.
			$ 
		\item (Restricted eigenvalue condition)
		There are  positive constants $\overline{c},\underline{c}$ and a sequence $m_N\rightarrow \infty $ such that, with probability approaching one, 
		$$		
		\underline{c}\leq \inf_{\beta \neq 0,\|\beta \|_{0}\leq m_N}\frac{\left(\frac{1}{N}\sum_{i=1}^N(b(X_i)^\top\beta)^2\right)^{1/2}}{\|\beta \|_{2}}\leq \sup_{\beta\neq 0,\|\beta \|_{0}\leq m_N}\frac{\left(\frac{1}{N}\sum_{i=1}^N(b(X_i)^\top\beta)^2\right)^{1/2}}{\|\beta \|_{2}}\leq \overline{c},
		$$
		$$
		\underline{c}\leq \inf_{\beta \neq 0,\|\beta \|_{0}\leq m_N}\frac{\left(\frac{1}{N}\sum_{i=1}^N(\frac{\partial}{\partial x_1} b(X_i)^\top\beta)^2\right)^{1/2}}{\|\beta \|_{2}}\leq \sup_{\beta\neq 0,\|\beta \|_{0}\leq m_N}\frac{\left(\frac{1}{N}\sum_{i=1}^N(\frac{\partial}{\partial x_1} b(X_i)^\top\beta)^2\right)^{1/2}}{\|\beta \|_{2}}\leq \overline{c},
		$$
$$		\sup_{\beta\neq 0,\|\beta \|_{0}\leq m_N}\left|\frac{\frac{1}{N}\sum_{i=1}^N(b(X_i)^\top\beta)^2}{\int(b(x)^\top\beta)^2dF_X(x)}-1\right| + \sup_{\beta\neq 0,\|\beta \|_{0}\leq m_N}\left|\frac{\frac{1}{N}\sum_{i=1}^N(\frac{\partial}{\partial x_1} b(X_i)^\top\beta)^2}{\int(\frac{\partial}{\partial x_1}b(x)^\top\beta)^2dF_X(x)}-1\right| = o_P(1).
$$
		\item (Sparsity) $\sup_{q \in\mathcal{Q}^\delta}\|\beta_q\|_0  \leq s_b$  
		for a sequence $s_b$ satisfying 
		$s_b=o(m_N)$ and $\zeta_N s_b\sqrt{{\log(p_b)}/{N}} = o(1)$,
		where
		$
		\zeta_N
		= \sup_{x \in \mathcal{X}}\max_{j =1,\ldots,p_b}\max\left\{\left|b_j(x)\right|,\left|\frac{\partial}{\partial x_1}b_j(x)\right|\right\}.
		$ 
		\item (Approximation error)
		$$
		\sup_{q \in \mathcal{Q}^\delta}
		\left(\int\left(\frac{\partial}{\partial x_1}(m_0(x,q)-\Lambda(b(x)^\top\beta_{q}))\right)^2dF_X(x)\right)^{1/2}
		 = O(\sqrt{s_b \log(p_b)/N})$$
		$$\sup_{x \in \mathcal{X},q\in \mathcal{Q}^\delta}\left|\frac{\partial}{\partial x_1}(m_0(x,q)-\Lambda(b(x)^\top\beta_{q}))\right|=O(\zeta_N s_b\sqrt{{\log(p_b)}/{N}}).
		$$
		\end{enumerate}
	\label{assn:Lasso}
\end{assumption}

Several remarks are in order. First, Assumption \ref{assn:Lasso}.1 is the common regularity condition. Second, Assumptions \ref{assn:Lasso}.2--\ref{assn:Lasso}.4 are common in the literature of logistic regressions with an $\ell_1$ penalty. See, for instance, \cite{BCFH13}.  Third, the various bounds for $m_1(x,q)$, $\beta_{q}$, $(m_0(X,q)-\Lambda(b(X)^\top\beta_{q}))$ need to hold uniformly over $q \in \mathcal{Q}^\delta$ because the estimator $\{\hat{q}_\tau: \tau \in \Upsilon\}$ belongs to $\mathcal{Q}^\delta$ for any fixed $\delta$ with probability approaching one. Fourth, as formally stated in Theorem \ref{thm:mhat} in the Online Supplement, Assumption \ref{assn:Lasso} can bound the estimation error for the logistic Lasso estimation: 
$$
\sup_{q \in \mathcal{Q}^\delta}\int\left|\hat{m}_{j}(x,q)-{m}_j(x,q)\right|^2dF_X(x) = O_P\left(\frac{s_b \log(p_b)}{N}\right)
$$
and 
$$
\sup_{q \in \mathcal{Q}^\delta,x \in \mathcal{X}}\left| \hat{m}_{j}(x,q) - m_j(x,q)\right|= O_P\left( \zeta_N s_b\sqrt{\frac{\log(p_b)}{N}}\right).
$$

We provide the regularity condition for the Riesz representer estimation of $\omega(x)$.
\begin{assumption}
	\begin{enumerate}
		\item[]
		\item (Boundedness) There is a constant $C$ such that $\max_{1\leq j \leq p_h}|h_j(X)| \leq C$ with probability one.
		\item (Estimation error) $\|\hat{G}- G\|_\infty + \|\hat{M}- M\|_\infty = O_P\left(\sqrt{\frac{\log(p_h)}{N}}\right)$.
		\item (Sparsity) Let  $s_h =C \left( \frac{\log(p_h)}{N}\right)^{-1/(1+2\xi)}$ for $C>1$, $\xi\geq 1/2$. Then, there is $\overline{\rho}$ with $\|\overline{\rho}\|_0 \leq s_h$ such that 
		$$
		\left(\int (\omega(x) - h(x)^\top \overline{\rho})^2dF_X(x)\right)^{1/2} \leq C (s_h)^{-\xi} \quad \text{and} \quad \sup_{x \in \mathcal{X}}|\omega(x)-h(x)^\top \overline{\rho}|=o(1).
		$$
		\item (Restricted eigenvalue condition) $G$ and $\hat{G}$'s eigenvalues are uniformly bounded in $n$, with probability approaching one. Also, 	there are  positive constants $\overline{c},\underline{c}$ and $m_N$ with $s_h = o(m_N)$ such that, with probability approaching one, 
$$
\underline{c}\leq \inf_{\rho \neq 0,\|\rho \|_{0}\leq m_N}\frac{\rho^\top \hat{G} \rho}{\|\rho\|_{2}^2}\leq \sup_{\Delta \neq 0,\|\rho \|_{0}\leq m_N}\frac{\rho^\top \hat{G} \rho}{\|\rho\|_{2}^2}\leq \overline{c} \quad \text{and}
$$
$$\underline{c}\leq \inf_{\rho \neq 0,\|\rho \|_{0}\leq m_N}\frac{\rho^\top G \rho}{\|\rho\|_{2}^2}\leq \sup_{\rho \neq 0,\|\rho \|_{0}\leq m_N}\frac{\rho^\top G \rho}{\|\rho\|_{2}^2}\leq \overline{c}.
$$
\item (Tuning parameter and dimensionality of $h(X)$) $\sqrt{ \log(p_h)/N}=o(\lambda_{R})$ and $\lambda_{R}=o(N^c\sqrt{\log(p_h)/N})$ for every $c>0$, and $p_h \leq C N^C$ for some $C>0$.	
	\end{enumerate}
	\label{ass:rr}  
\end{assumption} 

Assumption \ref{ass:rr} follows \cite{chernozhukov2018automatic} to which we refer readers for more discussion. Specifically, we have their $\varepsilon_n=\sqrt{ \log(p_h)/N}$ and $r = \lambda_{R}$ and Assumption \ref{ass:rr}.4 implies \citet[Assumption 3]{chernozhukov2018automatic} by \citet[Lemma 4.1]{BRT09}. 
Theorem \ref{thm:omegahat} in the Online Supplement shows that Assumption \ref{ass:rr} can bound the estimation error for $\hat\omega(x)$:  
$$
\int \left(\hat\omega(x) - \omega(x)\right)^2dF_X(x) = o_P(N^c s_h \log(p_h)/N)
\mbox{ and }
\sup_{x \in \mathcal{X}}|\hat\omega(x) - \omega(x)|  = o_P(1).
$$
for all $c>0$. In Theorem \ref{thm:omegahat}, we also show the Riesz representer $\hat\omega(x)$ belongs to a class of functions whose entropy or complexity level is well-controlled. Such a result is new to the literature and essential for our theory as we use all the observations to estimate $\omega(x)$, and thus, are subject to the model selection bias. \cite{chernozhukov2018automatic} circumvent such bias via cross-fitting. In the Online Supplement, we also consider cross-fitting which can accommodate complicated general machine learning estimators for $\omega(x)$.

\subsection{Testing $UQPE(\tau)=0, \forall\tau \in \Upsilon$}\label{sec:testing_uqpe}
As mentioned earlier, testing $UQPE(\tau)=0, \forall \tau \in \Upsilon$ is equivalent to testing $\theta(\tau)=0, \forall\tau \in \Upsilon$. We can reject the null hypothesis if the constant zero function over $\Upsilon$ does not belong to $CB^\theta_\Upsilon$. In this section, we show that the proposed confidence band $\text{CB}^\theta_\Upsilon$ covers the true $\theta(\tau)$ uniformly over $\tau \in \Upsilon$ with the correct asymptotic size.

We impose an additional rate condition about upper bounds on $s_h$ and $s_b$. 
\begin{assumption}\label{assn:rate_combined1}
$(s_b\log(p_b)+s_h\log(p_h))^2 = o(N^{\frac{d}{2+d}})$, and there is some $c>0$ such that $\pi_N^2(s_h \log(p_h) + s_b \log(p_b)) = o(1)$ where $\pi_N=\sqrt{N^{2c} s_h\log(p_h)/N + (\zeta_N^{4/(2+d)}s_b^{(4+d)/(2+d)})\log(p_b)/N}$.
\end{assumption}

When $\omega(\cdot)$ defined in Assumption \ref{assn:identification} is bounded so that $d = \infty$ and $\zeta_N$ defined in Assumption \ref{assn:Lasso} is also bounded, $\pi_N$ is roughly equal to $\sqrt{N^{2c}s_h \log(p_h)/N + s_b \log(p_b)/N}$, which is just the convergence rate for the first-stage estimators. In this case, this additional condition holds as long as $s_h \log(p_h) + s_b \log(p_b)= o(N^{1/2-c})$ for some $c>0$, which implies 
$$
\sqrt{\frac{s_h \log (p_h)}{N}} + \sqrt{\frac{s_b \log (p_b)}{N}}  = o(N^{-1/4}). 
$$
It means the convergence rate of the nuisance functions should be faster than the rate of $N^{1/4}$. Such a rate is sufficient for the influence function representation for $\hat{ \theta}(\tau)$ and $\hat{ \theta}^*(\tau)$ (in Theorem \ref{thm:theta}) due to the use of doubly/locally robust moment.

Theorem \ref{cor:theta} provides a sufficient condition for the correct asymptotic size of  the proposed confidence band $\text{CB}^\theta_\Upsilon$. 
It follows as a corollary of Theorem \ref{thm:theta} in the Online Supplement. 

\begin{theorem}\label{cor:theta}
Suppose $\sup_{\tau \in \Upsilon}\left|\sqrt{N}\hat{ \sigma}^\theta(\tau) - \sqrt{Var(\mathrm{IF}_i^\theta(\tau))}\right| = o_P(1)$ with $\mathrm{IF}_i^\theta(\tau) = m_{1}(X_i,{q}_\tau) - \omega(X_i)(1\{Y_i \leq {q}_\tau\} - m_{0}(X_i,{q}_\tau))-\theta(\tau)+ \frac{\frac{\partial}{\partial q} \mathbb{E}m_{1}(X,{q}_\tau)}{f_Y({q}_\tau)}(\tau - 1\{Y_i \leq {q}_\tau\})$. If Assumptions \ref{assn:identification}--\ref{assn:rate_combined1} hold,  then $$\mathbb{P}\left(\{\theta(\tau):\tau \in \Upsilon\} \in CB^\theta_\Upsilon\right) \rightarrow 1-\alpha.$$
\end{theorem}

\subsection{Confidence Band for $\{UQPE(\tau): \tau \in \Upsilon\}$}

In this section, we consider the confidence band for  $\{UQPE(\tau): \tau \in \Upsilon\}$, which can be used to infer the entire trajectory of $UQPE(\tau)$ over $\tau \in \Upsilon$. Recall that we use the kernel function $K_1(\cdot)$ in the kernel density estimation of $f_Y(\cdot)$. We impose the following conditions for the kernel function $K_1(\cdot)$ and the bandwidth parameter $h_1$. 

\begin{assumption}\label{assn:fkernel}
1. $K_1(\cdot)$ is a second-order symmetric kernel function with a compact support. 
2. $h_1=c_1N^{-H}$ for some positive constant $c_1$ and some $1/2>H \geq 1/5$.
\end{assumption}

We impose an additional rate condition about upper bounds on $s_h$ and $s_b$. 
\begin{assumption}\label{assn:rate_combined2}
$\log(N)h_1(s_b\log(p_b)+s_h\log(p_h))^2 = o(N^{\frac{d}{2+d}})$ and 
there is some $c>0$ such that $ \log(N)h_1\pi_N^2(s_h \log(p_h) + s_b \log(p_b)) = o(1)$ where $\pi_N$ is defined in Assumption \ref{assn:rate_combined1}.
\end{assumption}      
This assumption is weaker than Assumption \ref{assn:rate_combined2}, as $\log(N)h_1=o(1)$. In other words, if $d$ is sufficiently large and  $\zeta_N$ is bounded, such a condition holds as long as $\sqrt{\log(N)h_1}(s_h \log(p_h) + s_b \log(p_b))= o(N^{1/2-c})$ for some $c>0$.
      
The following theorem summarizes the validity for the bootstrap inference. 
The main takeaway from this theorem is that, by using the doubly robust method, we greatly relax the requirements on the sparsity and the number of effective covariates for $m_0(x,q)$ and $m_1(x,q)$, at the cost of imposing a sparsity condition on $\omega(x)$.\footnote{The leading term of the score function is $\frac{\theta(\tau)}{f_Y^2({q}_\tau)h_1}K_1\left(\frac{Y_i - {q}_\tau}{h_1}\right)$, but it does not imply that the doubly robust estimation for $\theta(\tau)$ is unnecessary. In fact, $\sup_{\tau \in \Upsilon}|\hat{ \theta}(\tau) - \theta(\tau)|$ is asymptotically negligible compared to $\frac{\theta(\tau)}{f_Y^2({q}_\tau)h_1}K_1\left(\frac{Y_i - {q}_\tau}{h_1}\right)$ if $\pi_N(\sqrt{s_h \log(p_h)/N} + \sqrt{s_b \log(p_b)/N}) + N^{-(1+d)/(2+d)}(s_h \log(p_h) + s_b \log(p_b))=o((\log(N)Nh_1)^{-1/2})$. If $h_1 = N^{-1/5}$, $d$ is sufficiently large, $c$ is arbitrarily small, and  $\zeta_N$ is polylogarithmic, such a condition holds if $s_h \log(p_h) + s_b \log(p_b)= o(N^{3/5})$ up to some  polylogarithmic factor.  On the other hand, if we do not use the doubly robust method, the estimation error of $\theta(\tau)$ is $\sqrt{s_b\log(p_b)/N}$, which is asymptotically negligible if $\sqrt{s_b\log(p_b)/N} = o((\log(N)Nh_1)^{-1/2})$. Such a condition would require $s_h \log(p_h) = o(N^{1/5})$ up to some  polylogarithmic factor.}

\begin{theorem}\label{cor:infer}
Suppose $\sqrt{Nh_1}=o(h_1^{-2})$, $h_1Var(\mathrm{IF}_i(\tau))$ is bounded away from zero, and $\sup_{\tau \in \Upsilon} \left\vert \sqrt{Nh_1}\hat\sigma(\tau)-\sqrt{h_1Var(\mathrm{IF}_i(\tau))}\right\vert=o_P(\log^{-1/2}(N))$, where $\mathrm{IF}_i(\tau) = \frac{\theta(\tau)}{f_Y^2({q}_\tau)h_1}K_1\left(\frac{Y_i - {q}_\tau}{h_1}\right)$. 
If Assumptions \ref{assn:identification}--\ref{ass:rr}, \ref{assn:fkernel},  \ref{assn:rate_combined2} hold, then
$$
\mathbb{P}(\{UQPE(\tau):\tau \in \Upsilon\}\in CB_\Upsilon)\rightarrow 1-\alpha.
$$
\end{theorem}

Theorem \ref{cor:infer} is a direct consequence of the linear expansions for $\widehat{UQPE}(\tau)$ and $\widehat{UQPE}^*(\tau)$ (formally stated in Theorem \ref{thm:UQPE} in the Online Supplement) and the strong approximation theory developed by  \cite{CCK14-anti,CCK14}. To compute $\hat\sigma(\tau)$, we can use either the plug-in method or the bootstrap method. For these methods, the convergence rate of $\sqrt{Nh_1}\hat\sigma(\tau)$ is polynomial in $N$, which implies $o_P(\log^{-1/2}(N))$.

\section{Simulation Studies}\label{sec:simulation}

In this section, we use Monte Carlo simulations to study the finite sample performance of the proposed method of estimation and inference for the UQPE.

Consider the following set of alternative data-generating designs. The outcome variable is generated according to the partial linear high-dimensional model
$$
Y\mid X \sim N\left(g(X_1) + \sum_{j=2}^p \alpha_j X_j, \ \ 1\right),
$$
where the function $g(x)$ is defined in the following three ways: $g(x) =  x$ in DGP 1, 
$g(x) =  x - 0.10 \cdot x^2$ in DGP 2, and $g(x) =  x - 0.10 \cdot x^2 + 0.01 \cdot x^3$ in DGP 3. The high-dimensional controls $(X_1,...,X_p)^\top$ are generated by
$$
X_1\mid (X_2,...,X_p) \sim N\left(\sum_{j=2}^p \gamma_j X_j ,\ 1\right)
\mbox{\ \ and\ \ }(X_2,...,X_p) \sim N(0,\Sigma_{p-1}),
$$
where  $\Sigma_{p-1}$ is the $(p-1) \times (p-1)$ variance-covariance matrix  whose $(r,c)$-element is $0.5^{2(|r-c|+1)}$. Note that this data-generating process induces dependence of the control $X_1$ of main interest on the rest of the $p-1$ controls $(X_2,\ldots,X_p)^\top$, as well as the dependence among the $p-1$ controls $(X_2,\ldots,X_p)^\top$. For the high-dimensional parameter vectors in the above data-generating model, we consider the following four cases of varying sparsity levels:
\begin{align*}
\text{(i)}
& \quad
(\alpha_2,\ldots,\alpha_p)^\top=(\gamma_2,\ldots,\gamma_p)^\top
=
(0.5^2,0.5^3,...,0.5^p)^\top,
\\
\text{(ii)}
& \quad
(\alpha_2,\ldots,\alpha_p)^\top=(\gamma_2,\ldots,\gamma_p)^\top
=
(0.5^2,0.5^{5/2},...,0.5^{(p+2)/2})^\top,
\\
\text{(iii)}
& \quad
(\alpha_2,\ldots,\alpha_p)^\top=(\gamma_2,\ldots,\gamma_p)^\top
=
(0.5^2,0.5^{7/3},...,0.5^{(p+4)/3})^\top,
\qquad\text{and}\\
\text{(iv)}
& \quad
(\alpha_2,\ldots,\alpha_p)^\top=(\gamma_2,\ldots,\gamma_p)^\top
=
(0.5^2,0.5^{9/4},...,0.5^{(p+6)/4})^\top.
\end{align*}

We follow the general estimation and inference approach outlined in Section \ref{sec:scoreconstruction}. We set $h(x) = (x^\top,(x^2)^\top,(x^3)^\top)^\top$ for estimation of $\omega(x)$, and 
set $b(x) = (x^\top,(x^2)^\top,(x^3)^\top)^\top$ for estimation of $m_0$ and $m_1$.
For the choice of $h_1$, we under-smooth the rule-of-thumb optimal choice as $h_1 = 1.06\sigma(Y) N^{-1/5-0.01}$. For each design, we use 500 iterations of Monte Carlo simulations to compute the mean, bias, and root mean square error (RMSE) of the estimate, as well as the 95\% uniform coverage over the set $[0.20,0.80]$ of quantiles. To evaluate the bias, RMSE, and the 95\% uniform coverage, we first numerically approximate the true UQPE by large-sample Monte Carlo simulations. Across sets of Monte Carlo simulations, we vary the DGP $\in \{\text{DGP 1, DGP 2, DGP 3}\}$ and the sparsity design $\in \{\text{(i)},\text{(ii)},\text{(iii)},\text{(iv)}\}$, while we fix the sample size $N  = 500$ and the dimension $p = 100$ throughout.

Table \ref{tab:simulation_approximate_sparsity} summarizes the simulation results under the sparsity designs (i) and (ii). We can make the following three observations in these results. First, the bias of our UQPE estimator is small, especially relative to the RMSE. This feature of the results supports the fact that our estimator mitigates the bias via the use of the doubly robust score. Second, the RMSE decreases as the sample size increases. Third, the 95\% uniform coverage frequencies are close to the nominal probability, namely, 0.95. This feature of the results supports our theory on the asymptotic validity of the bootstrap inference. From these simulation results, we confirm the main theoretical properties of the proposed method of estimation and inference for the UQPE across alternative data-generating processes. Table \ref{tab:simulation_approximate_sparsity_less_sparse} shows the simulation results under the less sparse designs (iii) and (iv). While the bias and RMSE are slightly bigger here than those in Table \ref{tab:simulation_approximate_sparsity}, the magnitudes of changes are modest. In addition to the simulation designs introduced above, we also experimented with other designs, and the simulation results are very similar and support the main theoretical properties of our proposed method as well -- see Appendix \ref{sec:additional_simulation} in the online supplement.

\begin{table}[tbh]
\centering
\scalebox{1}{
\begin{tabular}{ccccccccccccccc}
\multicolumn{14}{c}{(i) The Most Sparse Design -- with the Doubly Robust Score}\\
\hline\hline
      &          &     &     &     &        & True && \multicolumn{3}{c}{Estimates} && \multicolumn{2}{c}{95\% Cover}\\
\cline{9-11}\cline{13-14}
DGP &  & $N$ & $p$ & $\tau$ && UQPE && Mean & Bias & RMSE && Point & Unif. \\
\hline
\multirow{4}{*}{1 (i)} & \multirow{8}{*}{} & \multirow{4}{*}{500} & \multirow{4}{*}{100} &
     0.20 && 1.00 && 1.03 & 0.03 & 0.16 && 0.948 &\multirow{4}{*}{0.956}\\
&&&& 0.40 && 1.00 && 1.02 & 0.02 & 0.13 && 0.948 &\\
&&&& 0.60 && 1.00 && 1.03 & 0.03 & 0.14 && 0.954 &\\
&&&& 0.80 && 1.00 && 0.99 &-0.01 & 0.16 && 0.948 &\\
\hline
\multirow{4}{*}{2 (i)} & \multirow{8}{*}{} & \multirow{4}{*}{500} & \multirow{4}{*}{100} &
     0.20 && 1.12 && 1.14 & 0.02 & 0.18 && 0.952 &\multirow{4}{*}{0.956}\\
&&&& 0.40 && 1.03 && 1.05 & 0.02 & 0.13 && 0.946 &\\
&&&& 0.60 && 0.95 && 0.98 & 0.03 & 0.13 && 0.950 &\\
&&&& 0.80 && 0.87 && 0.88 & 0.00 & 0.15 && 0.950 &\\
\hline
\multirow{4}{*}{3 (i)} & \multirow{8}{*}{} & \multirow{4}{*}{500} & \multirow{4}{*}{100} &
     0.20 && 1.14 && 1.17 & 0.03 & 0.18 && 0.950 &\multirow{4}{*}{0.950}\\
&&&& 0.40 && 1.04 && 1.06 & 0.02 & 0.13 && 0.942 &\\
&&&& 0.60 && 0.97 && 1.00 & 0.03 & 0.13 && 0.944 &\\
&&&& 0.80 && 0.91 && 0.90 & 0.00 & 0.13 && 0.952 &\\
\hline\hline
\\
\multicolumn{14}{c}{(ii) The Second Most Sparse Design -- with the Doubly Robust Score}\\
\hline\hline
      &          &     &     &     &        & True && \multicolumn{3}{c}{Estimates} && \multicolumn{2}{c}{95\% Cover}\\
\cline{9-11}\cline{13-14}
DGP &  & $N$ & $p$ & $\tau$ && UQPE && Mean & Bias & RMSE && Point & Unif. \\
\hline
\multirow{4}{*}{1 (ii)} & \multirow{8}{*}{} & \multirow{4}{*}{500} & \multirow{4}{*}{100} &
     0.20 && 1.00 && 1.04 & 0.05 & 0.17 && 0.930 &\multirow{4}{*}{0.962}\\
&&&& 0.40 && 1.00 && 1.04 & 0.04 & 0.14 && 0.954 &\\
&&&& 0.60 && 1.00 && 1.04 & 0.04 & 0.15 && 0.920 &\\
&&&& 0.80 && 1.00 && 1.02 & 0.02 & 0.16 && 0.944 &\\
\hline
\multirow{4}{*}{2 (ii)} & \multirow{8}{*}{} & \multirow{4}{*}{500} & \multirow{4}{*}{100} &
     0.20 && 1.12 && 1.16 & 0.04 & 0.19 && 0.944 &\multirow{4}{*}{0.954}\\
&&&& 0.40 && 1.03 && 1.07 & 0.04 & 0.14 && 0.938 &\\
&&&& 0.60 && 0.95 && 0.99 & 0.04 & 0.14 && 0.918 &\\
&&&& 0.80 && 0.87 && 0.90 & 0.02 & 0.14 && 0.954 &\\
\hline
\multirow{4}{*}{3 (ii)} & \multirow{8}{*}{} & \multirow{4}{*}{500} & \multirow{4}{*}{100} &
     0.20 && 1.14 && 1.18 & 0.04 & 0.19 && 0.938 &\multirow{4}{*}{0.960}\\
&&&& 0.40 && 1.05 && 1.09 & 0.04 & 0.14 && 0.932 &\\
&&&& 0.60 && 0.97 && 1.01 & 0.04 & 0.14 && 0.922 &\\
&&&& 0.80 && 0.90 && 0.93 & 0.02 & 0.14 && 0.946 &\\
\hline\hline
\end{tabular}
}
\caption{Monte Carlo simulation results for the sparsity designs (i) and (ii). The true UQPE is numerically computed. The 95\% coverage is uniform over the set $[0.20,0.80]$.}
\label{tab:simulation_approximate_sparsity}
\end{table}

\begin{table}[tbh]
\centering
\scalebox{1}{
\begin{tabular}{ccccccccccccccc}
\multicolumn{14}{c}{(iii) The Third Most Sparse Design -- with the Doubly Robust Score}\\
\hline\hline
      &          &     &     &     &        & True && \multicolumn{3}{c}{Estimates} && \multicolumn{2}{c}{95\% Cover}\\
\cline{9-11}\cline{13-14}
DGP &  & $N$ & $p$ & $\tau$ && UQPE && Mean & Bias & RMSE && Point & Unif. \\
\hline
\multirow{4}{*}{1 (iii)} & \multirow{8}{*}{} & \multirow{4}{*}{500} & \multirow{4}{*}{100} &
     0.20 && 1.00 && 1.06 & 0.06 & 0.19 && 0.940 &\multirow{4}{*}{0.962}\\
&&&& 0.40 && 1.00 && 1.06 & 0.05 & 0.14 && 0.948 &\\
&&&& 0.60 && 1.00 && 1.05 & 0.05 & 0.14 && 0.932 &\\
&&&& 0.80 && 1.00 && 1.04 & 0.03 & 0.17 && 0.936 &\\
\hline
\multirow{4}{*}{2 (iii)} & \multirow{8}{*}{} & \multirow{4}{*}{500} & \multirow{4}{*}{100} &
     0.20 && 1.12 && 1.17 & 0.05 & 0.20 && 0.936 &\multirow{4}{*}{0.964}\\
&&&& 0.40 && 1.03 && 1.09 & 0.06 & 0.15 && 0.946 &\\
&&&& 0.60 && 0.95 && 1.00 & 0.05 & 0.14 && 0.936 &\\
&&&& 0.80 && 0.87 && 0.91 & 0.04 & 0.15 && 0.920 &\\
\hline
\multirow{4}{*}{3 (iii)} & \multirow{8}{*}{} & \multirow{4}{*}{500} & \multirow{4}{*}{100} &
     0.20 && 1.15 && 1.20 & 0.04 & 0.20 && 0.942 &\multirow{4}{*}{0.966}\\
&&&& 0.40 && 1.05 && 1.10 & 0.06 & 0.15 && 0.936 &\\
&&&& 0.60 && 0.97 && 1.02 & 0.05 & 0.14 && 0.936 &\\
&&&& 0.80 && 0.90 && 0.94 & 0.04 & 0.15 && 0.934 &\\
\hline\hline
\\
\multicolumn{14}{c}{(iv) The Least Sparse Design -- with the Doubly Robust Score}\\
\hline\hline
      &          &     &     &     &        & True && \multicolumn{3}{c}{Estimates} && \multicolumn{2}{c}{95\% Cover}\\
\cline{9-11}\cline{13-14}
DGP &  & $N$ & $p$ & $\tau$ && UQPE && Mean & Bias & RMSE && Point & Unif. \\
\hline
\multirow{4}{*}{1 (iv)} & \multirow{8}{*}{} & \multirow{4}{*}{500} & \multirow{4}{*}{100} &
     0.20 && 1.00 && 1.07 & 0.07 & 0.18 && 0.936 &\multirow{4}{*}{0.970}\\
&&&& 0.40 && 1.00 && 1.07 & 0.07 & 0.15 && 0.928 &\\
&&&& 0.60 && 1.00 && 1.06 & 0.06 & 0.15 && 0.930 &\\
&&&& 0.80 && 1.00 && 1.05 & 0.05 & 0.17 && 0.948 &\\
\hline
\multirow{4}{*}{2 (iv)} & \multirow{8}{*}{} & \multirow{4}{*}{500} & \multirow{4}{*}{100} &
     0.20 && 1.13 && 1.19 & 0.06 & 0.21 && 0.944 &\multirow{4}{*}{0.966}\\
&&&& 0.40 && 1.03 && 1.10 & 0.07 & 0.15 && 0.924 &\\
&&&& 0.60 && 0.95 && 1.01 & 0.06 & 0.14 && 0.938 &\\
&&&& 0.80 && 0.87 && 0.93 & 0.05 & 0.15 && 0.934 &\\
\hline
\multirow{4}{*}{3 (iv)} & \multirow{8}{*}{} & \multirow{4}{*}{500} & \multirow{4}{*}{100} &
     0.20 && 1.15 && 1.22 & 0.06 & 0.20 && 0.944 &\multirow{4}{*}{0.968}\\
&&&& 0.40 && 1.04 && 1.12 & 0.08 & 0.16 && 0.918 &\\
&&&& 0.60 && 0.97 && 1.03 & 0.06 & 0.14 && 0.928 &\\
&&&& 0.80 && 0.90 && 0.96 & 0.06 & 0.16 && 0.950 &\\
\hline\hline
\end{tabular}
}
\caption{Monte Carlo simulation results for the sparsity designs (iii) and (iv). The true UQPE is numerically computed. The 95\% coverage is uniform over the set $[0.20,0.80]$.}
\label{tab:simulation_approximate_sparsity_less_sparse}
\end{table}

To highlight the value added by our proposed method to the existing literature, we also experiment with the RIF-Logit estimator from \cite{firpo2009unconditional} as a benchmark. Table \ref{tab:simulation_FFL} summarizes the simulation results based on the RIF-Logit for the sample size of $N=500$ under the most sparse design (i).
Observe that, as the dimension $p$ increases from 25 to 50, the finite sample performance substantially degrades in terms of all of the displayed statistics, namely the bias, RMSE, and (pointwise and uniform) 95\% coverage frequencies. In particular, the uniform coverage frequency drops to zero even for the dimension that is as small as $p=50$. With the same sample size of $N=500$, on the other hand, our proposed method produces accurate coverage frequencies as well as accurate estimates for the even larger dimension $p=100$ as presented in Tables \ref{tab:simulation_approximate_sparsity} and \ref{tab:simulation_approximate_sparsity_less_sparse}. This comparison sheds lights on favorable finite sample performance of our proposed method when there are high-dimensional controls, in comparison with the existing alternative method. With that said, we would like to remark that we pay the costs of additional assumptions for nuisance parameter estimation, and hence there are tradeoffs between the existing procedure \citep{firpo2009unconditional} and our proposed method.

\begin{table}[t]
\centering
\scalebox{1}{
\begin{tabular}{ccccccccccccccc}
\multicolumn{14}{c}{The Conventional RIF-Loit Estimator under the Most Sparse Design (i)}\\
\hline\hline
      &          &     &     &        && True && \multicolumn{3}{c}{Estimates} && \multicolumn{2}{c}{95\% Cover}\\
\cline{9-11}\cline{13-14}
DGP &  & $N$ & $p$ & $\tau$ && UQPE && Mean & Bias & RMSE && Point & Unif. \\
\hline
\multirow{8}{*}{1 (i)} & \multirow{8}{*}{} & \multirow{4}{*}{500} & \multirow{4}{*}{25} 
   & 0.20 && 1.00 && 1.05 & 0.05 & 0.17 && 0.872 &\multirow{4}{*}{0.902}\\
&&&& 0.40 && 1.00 && 1.03 & 0.03 & 0.13 && 0.892 &\\
&&&& 0.60 && 1.00 && 1.03 & 0.03 & 0.13 && 0.898 &\\
&&&& 0.80 && 1.00 && 1.05 & 0.05 & 0.17 && 0.886 &\\
\cline{3-14}
&& \multirow{4}{*}{500} & \multirow{4}{*}{50}
   & 0.20 && 1.00 && 0.18 &-0.82 & 1.32 && 0.008 &\multirow{4}{*}{0.000}\\
&&&& 0.40 && 1.00 && 1.33 & 0.33 & 0.63 && 0.500 &\\
&&&& 0.60 && 1.00 && 1.33 & 0.33 & 0.57 && 0.474 &\\
&&&& 0.80 && 1.00 && 0.15 &-0.85 & 1.06 && 0.010 &\\
\hline
\multirow{8}{*}{2 (i)} & \multirow{8}{*}{} & \multirow{4}{*}{500} & \multirow{4}{*}{25}
   & 0.20 && 1.12 && 1.19 & 0.07 & 0.20 && 0.876 &\multirow{4}{*}{0.906}\\
&&&& 0.40 && 1.03 && 1.06 & 0.04 & 0.14 && 0.884 &\\
&&&& 0.60 && 0.96 && 0.99 & 0.03 & 0.12 && 0.900 &\\
&&&& 0.80 && 0.88 && 0.92 & 0.04 & 0.15 && 0.878 &\\
\cline{3-14}
&& \multirow{4}{*}{500} & \multirow{4}{*}{50}
   & 0.20 && 1.12 && 0.08 &-1.03 & 1.28 && 0.006 &\multirow{4}{*}{0.000}\\
&&&& 0.40 && 1.03 && 1.36 & 0.34 & 0.58 && 0.464 &\\
&&&& 0.60 && 0.95 && 1.24 & 0.29 & 0.49 && 0.528 &\\
&&&& 0.80 && 0.87 && 0.34 &-0.53 & 1.04 && 0.020 &\\
\hline
\multirow{8}{*}{3 (i)} & \multirow{8}{*}{} & \multirow{4}{*}{500} & \multirow{4}{*}{25}
   & 0.20 && 1.14 && 1.22 & 0.07 & 0.21 && 0.886 &\multirow{4}{*}{0.912}\\
&&&& 0.40 && 1.04 && 1.08 & 0.04 & 0.14 && 0.892 &\\
&&&& 0.60 && 0.97 && 1.01 & 0.03 & 0.13 && 0.900 &\\
&&&& 0.80 && 0.90 && 0.95 & 0.04 & 0.15 && 0.874 &\\
\cline{3-14}
&& \multirow{4}{*}{500} & \multirow{4}{*}{50}
   & 0.20 && 1.14 && 0.04 &-1.10 & 1.15 && 0.006 &\multirow{4}{*}{0.000}\\
&&&& 0.40 && 1.04 && 1.41 & 0.37 & 0.64 && 0.444 &\\
&&&& 0.60 && 0.97 && 1.28 & 0.31 & 0.52 && 0.502 &\\
&&&& 0.80 && 0.90 && 0.26 &-0.65 & 0.98 && 0.016 &\\
\hline\hline
\end{tabular}
}
\caption{Monte Carlo simulation results based on the conventional RIF-Logit estimator under the sparsity design (i). The true UQPE is numerically computed. The 95\% coverage is uniform over the set $[0.20,0.80]$.}
\label{tab:simulation_FFL}
\end{table}

In addition to the RIF-Logit estimator, we also experiment with a RIF-Lasso-Logit estimator. This estimator has not been formally investigated in the literature to our knowledge, but it coincides with a non-orthogonalized version of our procedure and hence serves as a useful benchmark to evaluate the benefits of our proposed doubly robust score. Table \ref{tab:simulation_double_vs_nodouble} shows the simulation results based on this procedure without the doubly robust score (right panel) compared with the results of our proposed procedure with the doubly robust score (left panel) copied from Table \ref{tab:simulation_approximate_sparsity}. While the coverage frequencies for our proposed method achieves the nominal probability of 95\%, those for the counterpart without the doubly robust score fall short of 95\%. These results show that the method without the doubly robust score incurs larger size distortions and demonstrate that it is useful to employ the doubly robust score as we do for our proposed method.

Finally, we consider the pointwise and uniform tests for $UQPE(\tau) =0$ by testing $\theta(\tau) = 0$. The corresponding simulation results are collected in Appendix \ref{sec:additional_simulation} in the online supplement.
\begin{table}[tbh]
\centering
\scalebox{1}{
\begin{tabular}{ccccccccccccccc}
\multicolumn{14}{c}{(i) The Most Sparse Design}\\
\hline\hline
\multicolumn{7}{c}{With the Doubly Robust Score}&&\multicolumn{7}{c}{Without the Doubly Robust Score}\\
\cline{1-7}\cline{9-15}
&&&&& \multicolumn{2}{c}{95\% Cover} &&&&&&& \multicolumn{2}{c}{95\% Cover}\\
\cline{6-7}\cline{14-15}
DGP & $N$ & $p$ & $\tau$ && Point & Unif. && DGP & $N$ & $p$ & $\tau$ && Point & Unif.\\
\cline{1-7}\cline{9-15}
\multirow{4}{*}{1 (i)} & \multirow{4}{*}{500} & \multirow{4}{*}{100} &
    0.20 && 0.948 &\multirow{4}{*}{0.956} && \multirow{4}{*}{1 (i)} & \multirow{4}{*}{500} & \multirow{4}{*}{100}  & 0.20 && 0.930 &\multirow{4}{*}{0.912}\\
&&& 0.40 && 0.948 &                       &&                                                                     &&& 0.40 && 0.912\\
&&& 0.60 && 0.954 &                       &&                                                                     &&& 0.60 && 0.902\\
&&& 0.80 && 0.948 &                       &&                                                                     &&& 0.80 && 0.910\\
\cline{1-7}\cline{9-15}
\multirow{4}{*}{2 (i)} & \multirow{4}{*}{500} & \multirow{4}{*}{100} &
    0.20 && 0.952 &\multirow{4}{*}{0.956} && \multirow{4}{*}{2 (i)} & \multirow{4}{*}{500} & \multirow{4}{*}{100}  & 0.20 && 0.924 &\multirow{4}{*}{0.910}\\
&&& 0.40 && 0.946 &                       &&                                                                     &&& 0.40 && 0.908\\
&&& 0.60 && 0.950 &                       &&                                                                     &&& 0.60 && 0.906\\
&&& 0.80 && 0.950 &                       &&                                                                     &&& 0.80 && 0.916\\
\cline{1-7}\cline{9-15}
\multirow{4}{*}{3 (i)} & \multirow{4}{*}{500} & \multirow{4}{*}{100} &
    0.20 && 0.950 &\multirow{4}{*}{0.950} && \multirow{4}{*}{3 (i)} & \multirow{4}{*}{500} & \multirow{4}{*}{100}  & 0.20 && 0.932 &\multirow{4}{*}{0.914}\\
&&& 0.40 && 0.942 &                       &&                                                                     &&& 0.40 && 0.910\\
&&& 0.60 && 0.944 &                       &&                                                                     &&& 0.60 && 0.908\\
&&& 0.80 && 0.952 &                       &&                                                                     &&& 0.80 && 0.922\\
\hline\hline
\\
\multicolumn{14}{c}{(ii) The Second Most Sparse Design}\\
\hline\hline
\multicolumn{7}{c}{With the Doubly Robust Score}&&\multicolumn{7}{c}{Without the Doubly Robust Score}\\
\cline{1-7}\cline{9-15}
&&&&& \multicolumn{2}{c}{95\% Cover} &&&&&&& \multicolumn{2}{c}{95\% Cover}\\
\cline{6-7}\cline{14-15}
DGP & $N$ & $p$ & $\tau$ && Point & Unif. && DGP & $N$ & $p$ & $\tau$ && Point & Unif.\\
\cline{1-7}\cline{9-15}
\multirow{4}{*}{1 (ii)} & \multirow{4}{*}{500} & \multirow{4}{*}{100} &
    0.20 && 0.930 &\multirow{4}{*}{0.962} && \multirow{4}{*}{1 (ii)} & \multirow{4}{*}{500} & \multirow{4}{*}{100}  & 0.20 && 0.914 &\multirow{4}{*}{0.920}\\
&&& 0.40 && 0.954 &                       &&                                                                      &&& 0.40 && 0.924\\
&&& 0.60 && 0.920 &                       &&                                                                      &&& 0.60 && 0.882\\
&&& 0.80 && 0.944 &                       &&                                                                      &&& 0.80 && 0.914\\
\cline{1-7}\cline{9-15}
\multirow{4}{*}{2 (ii)} & \multirow{4}{*}{500} & \multirow{4}{*}{100} &
    0.20 && 0.944 &\multirow{4}{*}{0.954} && \multirow{4}{*}{2 (ii)} & \multirow{4}{*}{500} & \multirow{4}{*}{100}  & 0.20 && 0.922 &\multirow{4}{*}{0.910}\\
&&& 0.40 && 0.938 &                       &&                                                                      &&& 0.40 && 0.920\\
&&& 0.60 && 0.918 &                       &&                                                                      &&& 0.60 && 0.890\\
&&& 0.80 && 0.954 &                       &&                                                                      &&& 0.80 && 0.926\\
\cline{1-7}\cline{9-15}
\multirow{4}{*}{3 (ii)} & \multirow{4}{*}{500} & \multirow{4}{*}{100} &
    0.20 && 0.938 &\multirow{4}{*}{0.960} && \multirow{4}{*}{3 (ii)} & \multirow{4}{*}{500} & \multirow{4}{*}{100}  & 0.20 && 0.914 &\multirow{4}{*}{0.920}\\
&&& 0.40 && 0.932 &                       &&                                                                      &&& 0.40 && 0.912\\
&&& 0.60 && 0.922 &                       &&                                                                      &&& 0.60 && 0.892\\
&&& 0.80 && 0.946 &                       &&                                                                      &&& 0.80 && 0.908\\
\hline\hline
\end{tabular}
}
\caption{Monte Carlo simulation results for the sparsity designs (i) and (ii) with the doubly robust score (left) and without the doubly robust score (right). The true UQPE is numerically computed. The 95\% coverage is uniform over the set $[0.20,0.80]$.}
\label{tab:simulation_double_vs_nodouble}
\end{table}

\section{Heterogeneous Counterfactual Marginal Effects of Job Corps Training}\label{sec:empirics}

The UQPE identifies counterfactual effects that are heterogeneous across outcome levels $Y$. This feature of the UQPE is useful for evaluating economic policies designed to benefit targeted subpopulations of the economy that are identified in terms of economic outcomes such as wage and income. For instance, major job training programs are designed to benefit targeted subpopulations of individuals who are low wage earners, i.e., lower quantiles of $Y$. In redesigning a job training program, a policy maker may want to choose such changes in $X$ that particularly benefit these targeted  subpopulations (with potentially lower wages) rather than the others (with potentially higher wages). Therefore, it is important for the policy maker to understand heterogeneous outcome gains (e.g., wage increase) of alternative counterfactual changes in $X$ across different subpopulations characterized by the levels of $Y$. The UQPE provides solutions to this goal.

While a rich set of empirical findings have been reported about the treatment effects of Job Corps, an analysis of heterogeneous counterfactual effects is missing in the literature to the best of our knowledge, despite its relevance to designing effective program policies and schemes as emphasized in the previous paragraph. Applying our proposed method, we analyze heterogeneous counterfactual marginal effects of Job Corps training on labor outcomes in this section. Specifically, it is important to find whether higher (respectively, lower) potential earners would benefit more (respectively, less) from counterfactually extending the duration of the training program. Since the entrance interview in Job Corps provides some information regarding the human capital of prospective trainees, answers to these empirical questions may possibly help the program designers to devise more efficient policies and schemes for the training programs. As such, we are interested in heterogeneous counterfactual marginal effects of the duration of the exposure to the program, as a continuous treatment variable, on labor outcomes measured by hourly wages. 

We identify and estimate the counterfactual distributional change given a large set of observed controls by taking advantage of our machine-learning-based method. For the outcome variable, we consider hourly wages. For the continuous treatment variables, we consider two seemingly similar but different measures: the duration in days of participation in Job Corps and the duration in days of actually taking classes in Job Corps. As will be shown shortly, these two definitions lead to qualitatively different empirical findings. We use 42 observed controls (and their powers). Table \ref{tab:summary_statistics} shows the summary statistics of our data. Different sets of observations are missing across different variables, and hence we use the intersection of observations that are non-missing across all the variables in use for our analysis. After dropping the missing observations, we are left with $N=481$ when we define the duration of participation in Job Corps as the treatment, while we are left with $N=368$ when we define the duration of actually taking classes in Job Corps. Taking the intersection of these two samples, we use a subsample of size $N=347$. Note that the dimension of covariates is relatively large given this effective sample size, and hence high-dimensional econometric methods are indispensable in the current application.

\begin{table}
\centering
\scalebox{0.90}{
\begin{tabular}{llccccc}
\hline\hline
&& 25th       &        &      & 75th       & Non-\\
&& Percentile & Median & Mean & Percentile & Missing \\
\hline
Outcome 
$Y$ & Hourly wage 
            & 4.750 & 5.340 & 5.892 & 6.500 & 7606\\
\hline
Treatment 
$X_1$ & Days in Job Corps 
                & 54.0 & 129.0 & 153.4 & 237.0 & 4748\\
      & Days taking classes 
      & 41.0 & 91.0 & 120.2 & 179.0 & 4207\\
\hline
Controls 
$X_{-1}$ & Age
                  & 17.00 & 18.00 & 18.43 & 20.00 & 14653\\
          & Female 
          & 0.000 & 0.000 & 0.396 & 1.000 & 14653\\
          & White
          & 0.000 & 0.000 & 0.303 & 1.000 & 14327\\
          & Black 
          & 0.000 & 1.000 & 0.504 & 1.000 & 14327\\
          & Hispanic origin
          & 0.000 & 0.000 & 0.184 & 0.000 & 14288\\
          & Native language is English 
          & 1.000 & 1.000 & 0.855 & 1.000 & 14327\\
  & Years of education
& 9.00 & 10.00 & 10.24 & 11.00 & 14327\\
& Other job trainings
& 0.000 & 0.000 & 0.339 & 1.000 & 13500\\
& Mother's education
& 11.00 & 12.00 & 11.53 & 11.53 & 11599\\
& Mother worked
& 1.000 & 1.000 & 0.752 & 1.000 & 14223\\
& Father's education
& 11.00 & 12.00 & 11.50 & 12.00 & 8774\\
& Father worked
& 0.000 & 1.000 & 0.665 & 1.000 & 12906\\
& Received welfare
& 0.000 & 1.000 & 0.563 & 1.000 & 14327\\
& Head of household
& 0.000 & 0.000 & 0.123 & 0.000 & 14327\\
& Number of people in household
& 2.000 & 3.000 & 3.890 & 5.000 & 14327\\
& Married
& 0.000 & 0.000 & 0.021 & 0.000 & 14327\\
& Separated
& 0.000 & 0.000 & 0.017 & 0.000 & 14327\\
& Divorced
& 0.000 & 0.000 & 0.007 & 0.000 & 14327\\
& Living with spouse
& 0.000 & 0.000 & 0.014 & 0.000 & 14235\\
& Child
& 0.000 & 0.000 & 0.266 & 1.000 & 13500\\
& Number of children
& 0.000 & 0.000 & 0.347 & 0.000 & 13500\\
& Past work experience
& 0.000 & 1.000 & 0.648 & 1.000 & 14327\\
& Past hours of work per week
& 0.000 & 24.00 & 25.15 & 40.00 & 14299\\
& Past hourly wage
& 4.250 & 5.000 & 5.142 & 5.500 & 7884\\
& Expected wage after training
& 7.000 & 9.000 & 9.910 & 11.000 & 6561\\
& Public housing or subsidy
& 0.000 & 0.000 & 0.200 & 0.000 & 14327\\
& Own house
& 0.000 & 0.000 & 0.411 & 1.000 & 11457\\
& Have contributed to mortgage
& 0.000 & 0.000 & 0.255 & 1.000 & 13951\\
& Past AFDC
& 0.000 & 0.000 & 0.301 & 1.000 & 14327\\
& Past SSI or SSA
& 0.000 & 0.000 & 0.251 & 1.000 & 14327\\
& Past food stamps
& 0.000 & 0.000 & 0.438 & 1.000 & 14327\\
& Past family income $\ge$ \$12K
& 0.000 & 1.000 & 0.576 & 1.000 & 14327\\
& In good health
& 1.000 & 1.000 & 0.871 & 1.000 & 14327\\
& Physical or emotional problem
& 0.000 & 0.000 & 0.049 & 0.000 & 14327\\
& Smoke
& 0.000 & 1.000 & 0.537 & 1.000 & 14327\\
& Alcohol
& 0.000 & 1.000 & 0.584 & 1.000 & 14327\\
& Marijuana or hashish
& 0.000 & 0.000 & 0.369 & 1.000 & 14327\\
& Cocaine
& 0.000 & 0.000 & 0.033 & 0.000 & 14327\\
& Heroin/opium/methadone
& 0.000 & 0.000 & 0.012 & 0.000 & 14327\\
& LSD/peyote/psilocybin
& 0.000 & 0.000 & 0.055 & 0.000 & 14327\\
& Arrested
& 0.000 & 0.000 & 0.266 & 1.000 & 14327\\
& Number of times arrested
& 0.000 & 0.000 & 0.537 & 1.000 & 14218\\
\hline\hline
\end{tabular}
}
\caption{Summary statistics of data.}
\label{tab:summary_statistics}
\end{table}

Observe that our sample consists of high-dimensional controls and the sample size that results from the aforementioned sample selection is not sufficiently large for conventional econometric methods of estimation and inference for the UQPE.
We therefore use our proposed method of estimation and inference for the UQPE that can accommodate a large dimension of controls via the use of the doubly robust score.

Using the same computer program as the one used for simulation studies presented in Section \ref{sec:simulation}, we obtain estimates, pointwise 95\% confidence intervals, and uniform 95\% confidence bands for $UQPE(\tau)$ for $\tau \in [0.20,0.80]$.
Table \ref{tab:job_corps} summarizes the results.
The row groups (I) and (II) report results for days in Job Corps as the treatment variable, while the row groups (III) and (IV) report results for days of taking classes in Job Corps as the treatment variable.
The row groups (I) and (III) report results for the hourly wage as the outcome variable, while the row groups (II) and (IV) report results for the logarithm of the hourly wage as the outcome variable.

\begin{table}[t]
\centering
\scalebox{1.00}{
\begin{tabular}{rllccrrrr}
\hline\hline
& Outcome & Treatment & $\tau$ & $\widehat{UQPE}(\tau)$ & \multicolumn{2}{c}{Pointwise 95\% CI} & \multicolumn{2}{c}{Uniform 95\% CB}\\
\hline
(I) &
Hourly & Days in     & 0.2 & 1.16 & [0.79 & 1.54] & [0.30 & 2.03]\\
  &wage  & Job Corps & 0.4 & 1.95 & [1.52 & 2.39] & [0.94 & 2.97]\\
  &      &           & 0.6 & 1.60 & [0.26 & 2.94] & [0.11 & 3.09]\\
  &   &              & 0.8 & 4.56 & [2.96 & 6.16] &[-0.67 & 9.79]\\
\hline
(II) &
Log    & Days in     & 0.2 & 0.20 & [0.13 & 0.27] & [0.02 & 0.38]\\
  &hourly& Job Corps & 0.4 & 0.50 & [0.30 & 0.69] &[-0.12 & 1.11]\\
  &wage  &           & 0.6 & 0.12 &[-0.15 & 0.38] &[-0.20 & 0.43]\\
  &   &              & 0.8 & 0.66 & [0.37 & 0.96] &[-0.04 & 1.37]\\
\hline
(III) &
Hourly & Days in     & 0.2 & 2.69 & [0.08 & 5.30] &[-19.06 & 24.44]\\
  &wage  & Job Corps & 0.4 & 2.66 & [2.07 & 3.25] &[-0.48 & 5.80]\\
&      & classes     & 0.6 & 1.14 & [0.00 & 2.29] &[-0.58 & 2.87]\\
&      &             & 0.8 & 5.30 & [2.76 & 7.84] &[-5.64 & 16.25]\\\hline
(IV) &
Log    & Days in     & 0.2 & 0.46 & [0.01 & 0.90] &[-4.24 & 5.15]\\
  &hourly& Job Corps & 0.4 & 0.64 & [0.38 & 0.89] &[-1.38 & 2.65]\\
&wage  & classes     & 0.6 & 0.17 &[-0.22 & 0.55] &[-0.25 & 0.58]\\
&      &             & 0.8 & 0.77 & [0.42 & 1.13] &[-1.29 & 2.84]\\
\hline\hline
\end{tabular}
}
\caption{Heterogeneous counterfactual marginal effects of days in Job Corps using  $p=42$ controls. The displayed values are thousand times the original values for ease of reading. The row groups (I) and (II) report results for days in Job Corps as the treatment variable, while the row groups (III) and (IV) report results for days of taking classes in Job Corps as the treatment variable.
The row groups (I) and (III) report results for the hourly wage as the outcome variable, while the row groups (II) and (IV) report results for the logarithm of the hourly wage as the outcome variable.
The results are based on the sample size of $N=347$.}
\label{tab:job_corps}
\end{table}

Overall, the magnitudes of the estimates are consistent with those from prior studies, and we also obtain the following new findings.\footnote{In the row group (I) in Table \ref{tab:job_corps} for instance, the \textit{daily} marginal effects range from 0.0012 to 0.0046 dollars. This magnitude is consistent with the 0.22 difference in average hourly wages between the treatment and control groups \citep*[][Table 3]{SBM2008}, where the average number of days in Job Corps for the treated group is 153.4 (Table \ref{tab:summary_statistics}).} 
First, observe that the none of the uniform 95\% confidence bands are contained in the negative reals. These results indicate that the counterfactual marginal effects of our interest are significantly negative for none of the heterogeneous subpopulations.
We next look into the heterogeneity of these effects. Observe in row (I) that the uniform 95\% confidence band for $\tau=0.2$ is contained in the positive reals while the uniform 95\% confidence band for $\tau=0.8$ intersects with the zero. These results imply heterogeneous statistical significance across quantiles. Specifically, we predict significantly positive counterfactual effects for lower wage earners ($\tau=0.2$, $0.4$ and $0.6$) and insignificant effects for higher wage earners ($\tau=0.8$). On the other hand, the point estimate is smaller for $\tau=0.2$ than that for $\tau=0.8$ in row (I). The larger effects for the subpopulation of higher potential earners (i.e., higher quantiles) could simply result from the scale effect. Heterogeneity in causal effects across different quantiles often vanishes once we take the logarithm of the outcome variable.\footnote{The relationship ${\partial Q_{\tau}(F^\epsilon_{\log(Y)})}/{\partial \epsilon}\big|_{\epsilon=0} = {(\partial Q_{\tau}(F^\epsilon_{Y})/\partial \epsilon)}/{Q_{\tau}(F^\epsilon_{Y})}\big|_{\epsilon=0}$ implies that the sign of the level and the logarithm coincide at the population level, even though the signs of empirical estimates and statistical significance may not coincide as in our empirical results.}
Therefore, we next consider the row group (II), where the outcome variable is defined as the logarithm of the hourly wage. Notice that, even in this row group, we continue to observe the same qualitative pattern as that in the row group (I). Namely, the point estimate is smaller for $\tau=0.2$ than that for $\tau=0.8$, but the uniform 95\% confidence band for $\tau=0.2$ is contained in the positive reals while the uniform 95\% confidence band for $\tau=0.8$ intersects with the zero. These results provide policy makers of confidence that extending the duration of exposures to the Job Corps program will benefit potential lower wage earners.

Once we turn to row groups (III) and (IV), where the treatment variable is now defined as days of taking classes in Job Corps, we no longer observe the aforementioned pattern of heterogeneous counterfactual marginal effects, and the uniform 95\% confidence bands globally intersect with the zero. However, if we implement the test $UQPE(\tau)=0, \forall\tau\in [0.20,0.80]$ based on $\hat\theta(\tau)$ as in Section \ref{sec:testing_uqpe}, then we actually reject this hypothesis of uniformly zero counterfactual marginal effects with the 95\% confidence.

In summary, we obtain the following three new findings about counterfactual marginal effects of the duration of exposure to Job Corps training on the hourly wage.
First, the effects are significantly negative for none of the heterogeneous subpopulations under consideration, regardless of the definition of the treatment variable and the definition of the outcome variable.
Second, the counterfactual marginal effects of days in Job Corps are significant for the subpopulation of lower wage earners, while they are insignificant for higher wage earners.
This result holds robustly regardless of whether we define the outcome variable as the hourly wage or the logarithm of it.
Third, we fail to detect the aforementioned pattern of counterfactual marginal effects once we define the treatment variable as days of taking classes in Job Corps, while we still reject the hypothesis of uniformly zero counterfactual marginal effects.
These results contain the following policy implications.
Extending the duration of exposures the Job Corps training program will be effective especially for the targeted subpopulations of lower potential wage earners.
However, these benefits may come from sources other than the experience of taking classes in the Job Corps training program.

Finally, to get further insights about our proposed method, we conclude this section with discussions of more details about what was implemented in the black box to produce the results reported in Table \ref{tab:job_corps}.
When estimating the core functions $m_0(x,q_\tau)$ and $m_1(x,q_\tau)$ by the lasso logit, we in fact select different subvectors of $b(X)$ across different quantiles $\tau$.
Table \ref{tab:selection} shows which controls and/or their powers are selected for each $\tau \in \{0.20,0.40,0.60,0.80\}$.
While there are some controls that are common (such as the intercept and the number of people in household) across all $\tau$, the selections are by no means uniform across $\tau$.
A researcher does not \textit{ex ante} know which variables among many in the list should be included for each quantile $\tau$.
Including all the potentially relevant controls would incur non-trivial size distortions -- recall the simulation results shown in Table \ref{tab:simulation_FFL} in Section \ref{sec:simulation}.
By our proposed method of inference that achieves the nominal size, on the other hand, it is those shown in Table \ref{tab:selection} that were selected to be relevant for each $\tau \in \{0.20,0.40,0.60,0.80\}$ in producing the estimation results reported in Table \ref{tab:job_corps}.

\begin{table}[t]
\centering
\scalebox{1.00}{
\begin{tabular}{rcccc}
\hline\hline
$\tau$ & 0.2 & 0.4 & 0.6 & 0.8\\
\hline
& Intercept & Intercept & Intercept & Intercept \\
\cline{2-5}
&           &           &           & Married \\
\cline{2-5}
&           &           & Separated & Separated \\
\cline{2-5}
&           &           &           & Living with spouse \\
\cline{2-5}
&           &           & Education & Education \\
\cline{2-5}
& Number of people & Number of people & Number of people & Number of people \\
& in household & in household & in household & in household \\
\hline\hline
\end{tabular}
}
\caption{The list of variables selected by the lasso logit estimation of $m_0(x,q_\tau)$ and $m_1(x,q_\tau)$ for $\tau \in \{0.2,0.4,0.6,0.8\}$.}
\label{tab:selection}
\end{table}

\section{Conclusion}\label{sec:concl}

Counterfactual analyses often involve high-dimensional controls. 
On the other hand, existing methods of estimation and inference for heterogeneous counterfactual changes are not compatible with high-dimensional settings.
In this paper, we therefore propose a novel doubly/locally robust score for debiased estimation and inference for the UQPE as a measure of heterogeneous counterfactual marginal effects.
A concrete implementation procedure is provide for estimation and multiplier bootstrap inference.
The online supplement additionally presents a general class of estimation and inference procedures.
Asymptotic theories are presented to guarantee that the bootstrap method works for size control.
Simulation studies support our theoretical properties.
Applying the proposed method of estimation and inference to survey data of Job Corps, the largest training program for disadvantaged youth in the United States, we obtain the following two policy implications.
First, extending the duration of exposures the Job Corps training program will be effective especially for the targeted subpopulations of lower potential wage earners.
Second, these benefits may come from sources other than the experience of taking classes in the Job Corps training program.

\bibliography{mybib}

\end{document}